\documentclass[sigconf]{acmart}
\usepackage{tikz}
\usetikzlibrary{arrows.meta,positioning,fit,calc}
\usepackage{booktabs}
\usepackage{multirow}
\usepackage{tabularx}
\usepackage{array}
\usepackage{placeins}

\renewcommand\footnotetextcopyrightpermission[1]{}
\setcopyright{none}
\acmDOI{}
\acmISBN{}

\definecolor{figamber}{RGB}{214,138,0}
\definecolor{figamberfill}{RGB}{255,247,224}
\definecolor{figgreen}{RGB}{46,125,50}
\definecolor{figgreenfill}{RGB}{232,245,233}
\definecolor{figblue}{RGB}{21,101,192}
\definecolor{figbluefill}{RGB}{227,242,253}
\definecolor{figgray}{RGB}{110,116,124}
\definecolor{figgrayfill}{RGB}{245,246,247}
\definecolor{figred}{RGB}{183,28,28}

\usepackage{xcolor}

\newcounter{researchquestion}

\newcommand{\researchquestion}[2]{%
  \refstepcounter{researchquestion}%
  \label{#1}%
  \par\smallskip
  \begingroup
  \setlength{\fboxsep}{3.5pt}%
  \setlength{\fboxrule}{0.4pt}%
  \noindent
  \fcolorbox{black!35}{black!4}{%
    \parbox{\dimexpr\linewidth-2\fboxsep-2\fboxrule\relax}{%
      \small
      \textbf{RQ\theresearchquestion.}\enspace #2%
    }%
  }%
  \endgroup
  \par\smallskip
}

\begin{document}

\title{No Trace, No Claim: Two Contracts for Database Agents}
\subtitle{Lessons from TGMS}

\author{Xiaofei Zhang}
\affiliation{%
  \institution{University of Memphis}
  \city{Memphis}
\state{Tennessee}
\country{USA}
}
\email{xiaofei.zhang@memphis.edu}

\begin{abstract}
LLM agents can generate database operations and explain their results, but current interfaces often leave a gap between the generated plan, the conditions under which its results were produced, and the claims presented to a user. We argue that agent-facing data systems need two enforceable contracts. A \emph{plan contract} defines what an agent may execute and reference, while an \emph{evidence contract} records the belief state, completeness, and provenance under which a result supports a claim.

We instantiate these contracts in TGMS, a bi-temporal graph system in which an LLM plans over a fixed temporal operator interface. A static verifier checks plans before execution, and a claim verifier checks typed claims against content-addressed traces. Live model runs exposed two failures missed by input-only schemas and
value-only grounding: models referenced nonexistent result fields and
reported page-local counts as complete-result counts. Result-field checking turns the first into a repairable rejection. Completeness propagation detects the second in all 15 controlled cases and misses all 15 when disabled.

On the frozen CollegeMsg workload, TGMS reaches 0.408 typed-answer accuracy, compared with 0.064--0.284 for the evaluated baselines. On Bitcoin-OTC, direct SQL over the same bi-temporal store matches TGMS, showing that the fixed operator interface does not provide a universal accuracy advantage. On correction probes, TGMS and bi-temporal SQL answer historical-belief questions, while latest-state baselines cannot. Before claim gating, 21 of 220 emitted answers contain an unsupported gated claim; after gating, none of the 199 emitted answers does, at the cost of 21 fewer answers. These results show that the plan contract makes invalid plans rejectable and execution reproducible under recorded conditions, while the evidence contract makes gated claims faithful to their cited evidence. Neither contract guarantees correct interpretation of user intent.

\end{abstract}

\maketitle

\section{Two failures, one missing layer}
\label{sec:intro}

LLM agents are becoming an important class of database client. Recent
work studies agent-first architectures, LLM query planners, and
semantic operator
pipelines~\cite{liu2026agentfirst,urban2024caesura,wang2025aop}. Much
of this work asks how to obtain useful plans and execute agent
workloads efficiently. We study a complementary question: what must a
data system record and enforce so that both an agent's execution and
its resulting claims remain checkable? Two incidents from the development of TGMS~\cite{tgms2026}, a bi-temporal graph system, frame the problem.

\noindent\textbf{Example A: the belief state.}
On March~1, the database records that account $a$ first contacted
account $b$ on February~9. On March~10, an ingest error is discovered,
and the recorded contact time is corrected. An auditor later asks,
``What did the system believe on March~1?''
A latest-state store cannot answer this question after the correction.
The limitation is not the expressiveness of SQL or Cypher. The earlier
version was discarded when the record was overwritten. A system that
must reproduce past decisions therefore needs to preserve the belief
state under which each computation ran.

\noindent\textbf{Example B: the truncated page.}
A reachability operator returns its default page: 100 rows from a
343-row result. The metadata records
\texttt{rows\_total}=343 and \texttt{truncated=true}, but the 14B Qwen
model ignores both fields. It correctly counts the 100 returned rows
and reports ``100 reachable accounts.'' The count is correct for the
returned page but false as a claim about the complete database result.
Value grounding alone cannot detect this error because the reported
entities and the derived count are consistent with the visible page.
The failure arises from ignoring a condition attached to the result:
the page is incomplete.

\noindent\textbf{The missing layer.}
These incidents expose two missing controls in the agent-database
interface. Input schemas describe
what may enter an operation, but they do not constrain how a generated
plan reads the result (\S\ref{sec:lessons}). Query results expose
values, but they do not always expose belief state, completeness, and
provenance in a form that downstream checking must respect. We
therefore place one contract before execution and another after
execution.
\begin{quote}
\textbf{Position.} Agent-facing data systems need two enforceable
contracts. A \emph{plan contract} defines the registered operations,
legal references, typed arguments and result fields, grounded
identifiers, and resource limits. An \emph{evidence contract} records
result identity, belief state, completeness, approximation status, and
provenance. The first governs admissible execution. The second makes claim
support auditable.
\end{quote}

The contracts therefore cannot be enforced solely at the agent or
prompting layer. The data system determines which belief state was
read, whether a result is complete, and which operator produced it.
It must record these execution conditions and expose them in a
machine-checkable form. Middleware can use this information to
validate plans and claims, but it cannot recover a belief state that
the database discarded or infer completeness and provenance that the
database never reported.

\begin{table*}[t]
\centering
\scriptsize
\caption{Guarantee envelope of the two contracts.}
\label{tab:guarantees}
\setlength{\tabcolsep}{4pt}
\renewcommand{\arraystretch}{1.12}
\begin{tabularx}{\textwidth}{@{}
  >{\raggedright\arraybackslash}p{0.24\textwidth}
  >{\raggedright\arraybackslash}p{0.17\textwidth}
  >{\raggedright\arraybackslash}X
  >{\raggedright\arraybackslash}p{0.25\textwidth}
@{}}
\toprule
\textbf{Failure} & \textbf{Contract check} & \textbf{Ensures} &
\textbf{Does not ensure} \\
\midrule
Invalid result-field reference & Result schema & Rejected before execution &
Intended operator choice \\
\midrule
Ungrounded identifier & Grounding rule & Rejected before execution &
Correct entity resolution \\
\midrule
Complete-result claim over incomplete evidence & Completeness propagation &
Cannot be certified as complete & Complete result enumeration \\
\midrule
Unrecorded or mismatched execution state & Execution-state record &
Executed state is identifiable and checkable & Intended state selection \\
\midrule
Claim not entailed by cited results & Claim verification &
Withheld or marked unsupported & Correctness of unchecked claim types \\
\bottomrule
\end{tabularx}
\end{table*}

This paper makes three contributions. First, we define the plan and
evidence contracts and state their guarantee envelope
(\S\ref{sec:contracts}). Second, we instantiate them in
TGMS, a bi-temporal property-graph system with a fixed
operator interface, deterministic execution, and trace-checked answers
(\S\ref{sec:lessons}). Third, we evaluate the design on frozen
workloads, mechanism ablations, a same-information SQL baseline, and
independently written questions (\S\ref{sec:eval}). The bi-temporal
graph setting stresses both contracts because plans compose several
operations, records may be corrected, and partial results can change
the meaning of an answer.

\section{The two contracts}
\label{sec:contracts}

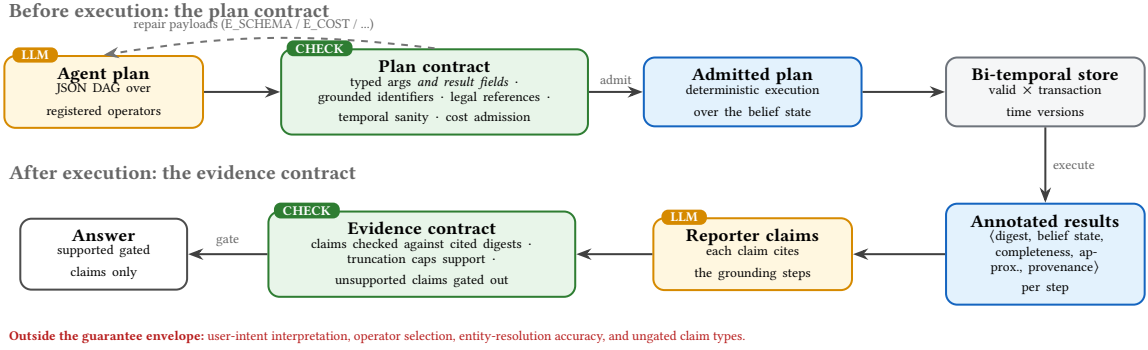
\begin{figure*}[t]
\centering
\resizebox{0.85\textwidth}{!}{%
\begin{tikzpicture}[
  every node/.style={font=\small},
  box/.style={rounded corners=5pt, align=center, inner ysep=4.5pt,
             inner xsep=6pt, line width=0.9pt},
  llm/.style={box, draw=figamber, fill=figamberfill},
  vrf/.style={box, draw=figgreen, fill=figgreenfill},
  exe/.style={box, draw=figblue, fill=figbluefill},
  gry/.style={box, draw=figgray, fill=figgrayfill},
  badge/.style={font=\scriptsize\bfseries, text=white,
                rounded corners=3pt, inner xsep=4pt, inner ysep=1.5pt},
  lbl/.style={font=\scriptsize, text=black!60},
  arr/.style={-{Stealth[length=2.4mm]}, line width=0.9pt, black!75}
]

\node[lbl, anchor=west] at (-0.2,1.25)
  {\normalsize\bfseries Before execution: the plan contract};
\node[llm, text width=27mm] (plan) at (1.4,0)
  {\textbf{Agent plan}\\[-1pt]
   {\scriptsize JSON DAG over\\registered operators}};
\node[badge, fill=figamber] at ($(plan.north west)+(0.5,0)$) {LLM};
\node[vrf, text width=44mm] (pc) at (6.6,0)
  {\textbf{Plan contract}\\[-1pt]
   {\scriptsize typed args \emph{and result fields} $\cdot$\\
    grounded identifiers $\cdot$ legal references $\cdot$\\[-1pt]
    temporal sanity $\cdot$ cost admission}};
\node[badge, fill=figgreen] at ($(pc.north west)+(0.62,0)$) {CHECK};
\node[exe, text width=30mm] (adm) at (11.6,0)
  {\textbf{Admitted plan}\\[-1pt]
   {\scriptsize deterministic execution\\over the belief state}};
\node[gry, text width=27mm] (sub) at (16.2,0)
  {\textbf{Bi-temporal store}\\[-1pt]
   {\scriptsize valid $\times$ transaction\\time versions}};
\draw[arr] (plan) -- (pc);
\draw[arr] (pc) -- node[lbl, above]{admit} (adm);
\draw[arr] (adm) -- (sub);
\draw[arr, dashed, black!55] (pc.north) to[bend right=8]
  node[lbl, above=0pt, pos=0.55]
  {repair payloads (E\_SCHEMA / E\_COST / ...)} (plan.north);

\node[lbl, anchor=west] at (-0.2,-1.25)
  {\normalsize\bfseries After execution: the evidence contract};
\node[exe, text width=27mm] (res) at (16.2,-2.55)
  {\textbf{Annotated results}\\[-1pt]
   {\scriptsize $\langle$digest, belief state,\\
    completeness, approx., provenance$\rangle$\\[-1pt] per step}};
\node[llm, text width=27mm] (rep) at (11.6,-2.55)
  {\textbf{Reporter claims}\\[-1pt]
   {\scriptsize each claim cites\\the grounding steps}};
\node[badge, fill=figamber] at ($(rep.north west)+(0.5,0)$) {LLM};
\node[vrf, text width=44mm] (ec) at (6.4,-2.55)
  {\textbf{Evidence contract}\\[-1pt]
   {\scriptsize claims checked against cited digests $\cdot$\\
    truncation caps support $\cdot$\\[-1pt]
    unsupported claims gated out}};
\node[badge, fill=figgreen] at ($(ec.north west)+(0.62,0)$) {CHECK};
\node[box, draw=black!70, text width=22mm] (out) at (1.4,-2.55)
  {\textbf{Answer}\\[-1pt]
   {\scriptsize supported gated\\claims only}};
\draw[arr] (sub.south) -- node[lbl, right]{execute} (res.north);
\draw[arr] (res) -- (rep);
\draw[arr] (rep) -- (ec);
\draw[arr] (ec) -- node[lbl, above]{gate} (out);

\node[lbl, text=figred, align=left, anchor=west] at (-0.2,-3.85)
  {\textbf{Outside the guarantee envelope:}
   user-intent interpretation, operator selection,
   entity-resolution accuracy, and ungated claim types.};
\end{tikzpicture}}
\vspace{-2ex}
\caption{The plan contract checks admissibility before execution.
Execution produces annotated results, and the evidence contract checks the
resulting claims. The contracts do not guarantee intent interpretation or
operator choice.}
\Description{A two-row loop. The upper row shows plan admission and execution
against a bi-temporal store. The lower row shows annotated results, reporter
claims, evidence checking, and a gated answer.}
\label{fig:contracts}
\end{figure*}

\noindent\textbf{Plan contract.}
A plan contract defines the admissible structure and references of an
agent-generated plan before execution. The plan must use a
machine-checkable representation in which each step invokes a
registered operation and references earlier results through explicitly
defined fields. Transformations that affect execution or downstream claims should
be represented as registered operations or otherwise made
machine-checkable.

Each registered operation declares its argument schema, result schema,
execution conditions, and resource model. The verifier checks the
plan's structure, argument types, data dependencies, result-field
references, grounded identifiers, relevant semantic constraints, and
resource bounds. In particular, a plan may reference only fields that
the producing operation declares. Deriving both the model-facing
interface and the verifier checks from the same registry keeps the
contract presented to the agent consistent with the contract enforced
by the system.

A rejected plan should receive structured repair information that
identifies the violated constraint and, when possible, the legal
alternatives or minimum change required for admission. Such feedback
does not relax the contract. It allows the agent to revise an
inadmissible plan while keeping every executed plan within the declared
interface and resource bounds.

\noindent\textbf{Evidence contract.}
An evidence contract defines the execution conditions under which a
result can support a downstream claim. Each plan step should therefore
produce not only a value, but also a machine-checkable evidence
descriptor that identifies the result and records the state under which
it was computed. The descriptor records the execution properties relevant to support,
including the applicable belief state, completeness, approximation
status, and provenance. These conditions must propagate through dependent operations. A result
derived from incomplete, stale, or approximate input should not
silently appear complete, current, or exact. The contract therefore
defines how evidence properties compose across a multi-step plan and
which forms of support remain valid after each transformation.


A downstream claim must identify the results on which it relies.
For claim types covered by the contract, a verifier checks whether
the cited evidence supports the claim under the recorded execution
conditions. Claims that do not satisfy the required proof obligation
cannot be presented as supported; an implementation may withhold them
or mark them explicitly as unsupported. When the interface emits only
supported claims, the operational rule is simple: no trace, no claim.
Example~B becomes detectable because a count derived from an
incomplete page cannot support a claim about the complete result, even
when the arithmetic over the visible rows is correct.

\noindent\textbf{Guarantee envelope.}
The two contracts establish three properties. First, \emph{plan
admissibility}: every executed plan conforms to the declared operation
interface, uses legal references and grounded identifiers, and remains
within its resource bounds. Second, \emph{execution reproducibility}: when operation semantics and execution states are versioned, and execution is deterministic, an
admitted plan can be re-executed under the same declared conditions to
reproduce its results. Third, \emph{claim
faithfulness}: for claim types covered by the evidence contract, every
emitted claim follows from its cited results under their recorded
execution conditions.

These properties do not imply \emph{task correctness}. An agent may
select an admissible plan that answers the wrong question and report
its result faithfully. The contracts constrain which plans may execute
and which claims their results may support; they do not determine what
the user intended or whether the agent selected the appropriate
operations. Table~\ref{tab:guarantees} summarizes this boundary.

\noindent\textbf{Belief state as execution evidence.}
The data state read by a computation is part of the meaning of its
result. When records can be corrected, the same operation may produce
different answers over different database versions. An evidence
contract must therefore identify the belief state under which each
result was produced and preserve that identity through downstream
computation. Bi-temporal storage is one way to preserve such
states~\cite{snodgrass1999}; snapshots, versioned logs, and other
state-preserving mechanisms may serve the same logical role.

\section{TGMS and two lessons from implementation}
\label{sec:lessons}

TGMS instantiates both contracts over a bi-temporal property-graph
store. An append-only event log preserves valid-time and
transaction-time versions of each update. Thirteen temporal operators
cover historical reads, snapshots and diffs, traversal, aggregation,
patterns, entity resolution, and deterministic computation. Each
operator is validated against an independent brute-force oracle on 500
randomized cases. The planner and reporter use open-weight models
served by vLLM~\cite{kwon2023vllm,qwen25}.

\noindent\textbf{Plan-contract implementation.}
The planner represents a task as a JSON DAG. Each step invokes one
operator from the registry and may reference declared fields produced
by earlier steps. References use a restricted path language and cannot
perform arithmetic; derived values are computed by a registered,
deterministic compute operator. For each operator, the registry declares its argument types, result
fields, temporal parameters, pagination behavior, and cost model. The
same declarations generate both the model-facing interface and the
verifier checks. Before execution, the verifier rejects malformed
DAGs, invalid argument types, cycles, illegal dependencies,
undeclared result-field references, ungrounded identifiers,
inconsistent temporal arguments, and plans that exceed the resource
budget. A rejection includes a structured error code and, when
possible, a concrete repair. For example, when a time-series request
exceeds the bucket cap, the verifier returns the minimum legal
stride rather than instructing the model to increase it.

\noindent\textbf{Evidence-contract implementation.}
The executor runs each admitted plan deterministically. It
canonically serializes each step result and records its SHA-256 digest.
Under the same store state, software version, and arguments,
re-execution reproduces both the result bytes and the digest. Each result also carries an
evidence descriptor containing its digest, transaction-time belief
state (\texttt{as\_of\_tt}), completeness information
(\texttt{rows\_total} and truncation flags), approximation status, and
producing operator. Conditions that affect interpretation propagate
to dependent steps; in particular, a result derived from truncated
input remains tainted by incompleteness. The reporter emits typed claims that cite the trace steps on which they
depend. For claim types covered by the verifier, TGMS checks whether
the cited results entail the claim under their recorded belief state
and completeness conditions. Unsupported claims are removed before
the answer is emitted. This realizes the rule \emph{no trace, no
claim}: a citation must identify an immutable result, and the result's
execution conditions must be sufficient for the stated claim.

The following incidents exposed gaps in the initial implementation
and motivated two checks in the final design.

\noindent\textbf{Incident 1: models invent result fields.}
In the first live probe run, every generated plan passed argument
validation, yet none executed successfully. The plans referenced
fields that the producing operators did not return, such as
\texttt{s2.count} instead of \texttt{s2.rows\_total}. Argument schemas
constrained values passed into an operator but did not constrain how
later steps read its result. We therefore added result schemas to the
registry and checked every inter-step reference before execution. On a three-task probe set constructed to reproduce the failure,
execution success increased from 0/3 to 3/3. In the 7B ablation
(Table~\ref{tab:ablations}), execution success is 0.55 with the check
and 0.23 without it. The failure is semantic rather than purely syntactic.
Grammar-constrained decoding~\cite{willard2023outlines} did not
address this mismatch and, in a separate 14B comparison, reduced both
first-plan acceptance and task accuracy. A syntactically valid
reference can still name a field that the producing operator does not
declare.

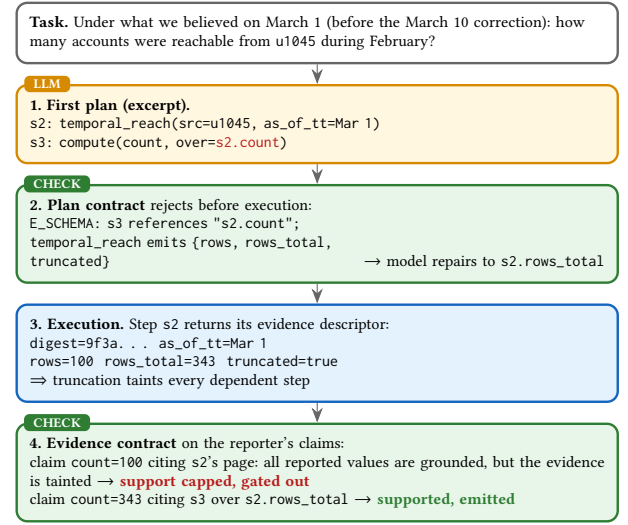
\begin{figure}[t]
\centering
\begin{tikzpicture}[
  stage/.style={rounded corners=4pt, align=left, inner sep=5pt,
                text width=76mm, line width=0.9pt, font=\scriptsize},
  tag/.style={font=\tiny\bfseries, text=white, rounded corners=2pt,
              inner xsep=3.5pt, inner ysep=1.5pt},
  node distance=2.6mm]
\node[stage, draw=black!60, fill=white] (task)
  {\textbf{Task.} Under what we believed on March~1 (before the
   March~10 correction): how many accounts were reachable from
   \texttt{u1045} during February?};
\node[stage, draw=figamber, fill=figamberfill, below=of task] (p1)
  {\textbf{1. First plan (excerpt).}\\
   \texttt{s2: temporal\_reach(src=u1045, as\_of\_tt=Mar 1)}\\
   \texttt{s3: compute(count, over=\textcolor{figred}{s2.count})}};
\node[tag, fill=figamber] at ($(p1.north west)+(0.42,0)$) {LLM};
\node[stage, draw=figgreen, fill=figgreenfill, below=of p1] (rej)
  {\textbf{2. Plan contract} rejects before execution:\\
   \texttt{E\_SCHEMA: s3 references "s2.count";}\\
   \texttt{temporal\_reach emits \{rows, rows\_total,}\\
   \texttt{truncated\}} \hfill $\to$ model repairs to
   \texttt{s2.rows\_total}};
\node[tag, fill=figgreen] at ($(rej.north west)+(0.55,0)$) {CHECK};
\node[stage, draw=figblue, fill=figbluefill, below=of rej] (ev)
  {\textbf{3. Execution.} Step \texttt{s2} returns its evidence
   descriptor:\\
   \texttt{digest=9f3a\ldots\ as\_of\_tt=Mar 1}\\
   \texttt{rows=100\ \ rows\_total=343\ \ truncated=true}\\
   $\Rightarrow$ truncation taints every dependent step};
\node[stage, draw=figgreen, fill=figgreenfill, below=of ev] (ver)
  {\textbf{4. Evidence contract} on the reporter's claims:\\
   claim \texttt{count=100} citing \texttt{s2}'s page: all reported values are
   grounded, but the evidence is tainted
   $\to$ \textcolor{figred}{\textbf{support capped, gated out}}\\
   claim \texttt{count=343} citing \texttt{s3} over
   \texttt{s2.rows\_total} $\to$
   \textcolor{figgreen}{\textbf{supported, emitted}}};
\node[tag, fill=figgreen] at ($(ver.north west)+(0.55,0)$) {CHECK};
\draw[-{Stealth[length=2mm]}, black!60] (task) -- (p1);
\draw[-{Stealth[length=2mm]}, black!60] (p1) -- (rej);
\draw[-{Stealth[length=2mm]}, black!60] (rej) -- (ev);
\draw[-{Stealth[length=2mm]}, black!60] (ev) -- (ver);
\end{tikzpicture}
\vspace{-1ex}\caption{One task through both contracts, assembled from the two
incidents of \S\ref{sec:lessons}: the invented result-field reference is rejected before execution;
the count over a truncated page survives value grounding but not the
evidence check.}
\Description{A four-stage example showing an invalid result-field reference rejected before execution, followed by completeness-aware claim checking after execution.}
\label{fig:worked}
\end{figure}

\noindent\textbf{Incident 2: a correct count over incomplete evidence.}
Example~B exposed a gap between value grounding and evidence
sufficiency. TGMS addresses it by recording result completeness,
propagating incompleteness through dependent steps, and incorporating
that condition into claim verification. A truncated page cannot
certify a claim about the complete result, even when arithmetic over
the visible rows is correct. Other
execution conditions require distinct treatment: \texttt{LIMIT} and
failed partitions may produce incomplete results, sampling produces
approximate results, and stale replicas produce results under an older
state. The evidence contract preserves these distinctions rather than
reducing support to a single Boolean
(\S\ref{sec:agenda}). Figure~\ref{fig:worked} shows how the two
incidents pass through the corresponding contracts in one end-to-end
task.

\FloatBarrier
\section{Evidence}
\label{sec:eval}
\setcounter{researchquestion}{0}
We evaluate four frozen workloads: CollegeMsg~\cite{panzarasa2009},
email-EU, a synthetic temporal-pattern workload, and the Bitcoin-OTC
signed trust network. Their test splits contain 94, 94, 102, and 94
tasks, respectively, and were fixed before evaluation. Gold answers are produced by the validated deterministic engine. The public artifact records the code,
configuration, and split identifiers for each result~\cite{tgms2026}. 
Because the benchmark and operator interface were co-designed,
accuracy measures planning and checking on questions expressible by
the current algebra. RQ~\ref{rq:coverage} separately evaluates coverage
on independently written questions. We report typed-answer accuracy:
counts and normalized entity sets require exact agreement, while
intervals require an intersection-over-union of at least 0.5. The
primary configuration uses Qwen2.5-14B-Instruct-AWQ at temperature~0;
CollegeMsg aggregates three seeds.

\begin{table}[t] \centering\small \caption{Frozen-test typed-answer accuracy. Bi-temporal SQL and TGMS use the same transaction history; the baselines right of the rule use latest-state representations.}\vspace{-2ex} \label{tab:main} \setlength{\tabcolsep}{3.3pt} \begin{tabular}{@{}lcc|ccc@{}} \toprule & TGMS & \begin{tabular}{@{}c@{}}bi-temp.\\SQL\end{tabular} & \begin{tabular}{@{}c@{}}text-to-\\Cypher\end{tabular} & \begin{tabular}{@{}c@{}}graph\\RAG\end{tabular} & \begin{tabular}{@{}c@{}}vector\\RAG\end{tabular} \\ \midrule CollegeMsg (94$\times$3) & \textbf{0.408} & 0.284 & 0.152 & 0.064 & 0.106 \\ email-EU (94) & \textbf{0.309} & 0.266 & 0.106 & 0.053 & -- \\ synthetic (102) & \textbf{0.314} & 0.265 & 0.157 & 0.029 & -- \\ Bitcoin-OTC (94) & 0.309 & \textbf{0.340} & 0.160 & 0.085 & -- \\ probes, CollegeMsg & \textbf{0.897} & 0.846 & 0.000 & 0.000 & 0.154 \\ probes, email-EU & \textbf{0.846} & \textbf{0.846} & 0.000 & 0.000 & -- \\ probes, Bitcoin-OTC & \textbf{0.923} & \textbf{0.923} & 0.000 & 0.000 & -- \\ \bottomrule \end{tabular} \end{table}

\researchquestion{rq:interface}{
What does the registered operator interface add beyond direct
queries over the same information?
}

The latest-state baselines lack transaction history, so their
performance on correction probes reflects an information limitation
rather than interface quality. We therefore add a same-information
baseline that gives the model direct SQL access to the same bi-temporal
DuckDB tables used by TGMS. The prompt documents the version schema and
temporal semantics, while the model, repair budget, execution limits,
and answer format remain fixed. This comparison changes the query
language, plan representation, and static checks together; it therefore
compares two interface designs rather than isolating one mechanism.

\FloatBarrier

The correction-probe rows in Table~\ref{tab:main} establish a
representation requirement.
On the correction probes, latest-state text-to-Cypher and graph RAG
cannot recover the earlier belief state because it is absent from
their inputs. Vector RAG succeeds only on some current-belief cases.
Transaction history is therefore a prerequisite for answering
historical-belief questions after corrections. Once this information gap is removed, TGMS and bi-temporal SQL perform
similarly on the correction probes. The fixed operator interface is
not what enables historical recovery; preserving and selecting the
appropriate belief state is.

The interface comparison on the full workloads is more nuanced.
TGMS significantly outperforms bi-temporal SQL on CollegeMsg
(difference $+0.124$, 95\% CI $[+0.071,+0.181]$), but the two are
statistically indistinguishable on Bitcoin-OTC (95\% CI
$[-0.117,+0.053]$). CollegeMsg contains more multi-step temporal
compositions, whereas many Bitcoin-OTC questions reduce to a single
query over a flatter schema. This pattern is consistent with a
registered operator interface helping on compositional workloads, but
the experiment does not isolate composition as the cause. The smaller
gains on email-EU and the synthetic workload are positive but
inconclusive because their confidence intervals include zero.

These results define the scope of the comparison. A fixed operator
interface does not provide a universal accuracy advantage, and the SQL
baseline does not implement the evidence contract; the experiment
therefore does not show that SQL results cannot be traced or checked.
A long-context vector-RAG control at $k{=}20$ used 32{,}475 tokens
per task and reached 0.021 accuracy, indicating that the retrieval
result is not explained solely by limited retrieval breadth.

\begin{table}[t]
\centering\small
\caption{Mechanism ablations on the CollegeMsg development split.
The truncation rows use a controlled generator and report detections.}\vspace{-2ex}
\label{tab:ablations}
\begin{tabular}{@{}llccc@{}}
\toprule
 & & \begin{tabular}{@{}c@{}}first-plan\\acceptance\end{tabular}
 & \begin{tabular}{@{}c@{}}execution\\success\end{tabular}
 & accuracy \\
\midrule
\multirow{2}{*}{7B}  & full contract & 0.23 & 0.55 & 0.136 \\
                     & no result-field check & 0.41 & 0.23 & 0.091 \\
\addlinespace[2pt]
\multirow{2}{*}{14B} & full contract & 0.50 & 0.65 & 0.409 \\
                     & no result-field check & 0.55 & -- & 0.318 \\
\midrule
\multicolumn{2}{@{}l}{truncated-count, taint on}
  & \multicolumn{3}{c}{15/15 (generator); 3 caught (natural split)} \\
\multicolumn{2}{@{}l}{truncated-count, taint off}
  & \multicolumn{3}{c}{0/15 (generator); 0 caught (natural split)} \\
\bottomrule
\end{tabular}
\end{table}

\researchquestion{rq:mechanisms}{
Which contract checks change execution and claim-support outcomes?
}

Table~\ref{tab:ablations} shows that removing result-field checking
raises first-plan acceptance but reduces execution success and final
accuracy. The verifier admits more plans only because it no longer
detects invalid inter-step references. Those errors then surface as runtime failures, after the structured
pre-execution repair opportunity has been lost. Result-field checking
therefore does not merely reject plans; it converts latent runtime
failures into structured pre-execution feedback that the repair loop
can act on. Completeness propagation has a similarly targeted effect. It detects
all truncated-count failures in the controlled generator, while
disabling it detects none. Only three verdicts change on the natural
development split because truncated results are uncommon there. The
controlled cases establish that propagation is necessary for this
failure class, while the natural split indicates how often the
mechanism is exercised in the observed workload.

\researchquestion{rq:gating}{
What does claim gating guarantee, and at what coverage cost?
}

An unsupported-answer rate is meaningful only together with answer
coverage: a system that emits nothing would otherwise appear perfectly
supported. Table~\ref{tab:coverage} therefore reports both quantities
on the frozen CollegeMsg campaign of 282 task runs. An answer is
\emph{emitted} if it contains at least one claim, and is
\emph{unsupported} if the verifier rejects at least one gated claim
about a count, value, entity, or ordering. The comparison isolates the effect of enforcement. In the ungated configuration, verifier verdicts are recorded but not
applied; 21 of 220 emitted answers contain an unsupported gated claim.
After gating, all gated claims in the 199 emitted answers pass the
verifier. The cost is 21 fewer emitted answers and a one-percentage-point
decrease in overall accuracy. Gating
therefore provides faithfulness for the claim types it covers by
trading coverage for support. It does not establish task correctness,
and temporal-pattern claims remain outside the current gating
envelope.


Table~\ref{tab:coverage} also shows that systems occupy different
coverage--accuracy tradeoffs. Bi-temporal SQL and text-to-Cypher emit more often than TGMS but
have lower conditional accuracy, while graph RAG and vector RAG emit
less often. TGMS abstains
when its repair loop cannot produce an admissible plan. This behavior
is most visible on the eight planted-pattern tasks, for which the
current pipeline produces no answer. Abstention avoids emitting an answer on those tasks, but it does
not solve gaps in planning or operator coverage.

Mutation tests evaluate the verifier independently of end-to-end
planning. It detects all 500 injected count and identifier faults,
all 60 values taken from the wrong belief state, all 100 real-but-uncited entities, and all tested truncation failures, with no false
positives on 87 clean answers.

Two additional mutations clarify the semantics of claim verification.
Changing a citation does not necessarily invalidate a claim: in 64 of
100 cases, the substituted step still entails it, and the verifier
correctly accepts the claim. It rejects the remaining 36 cases because
the replacement evidence is insufficient. By contrast, membership
checking cannot establish set completeness. The verifier confirms that
each reported entity occurs in the evidence, but detects none of 100
cases in which valid members are omitted. Complete-set claims therefore
require a stronger proof obligation, such as access to the full result
or a database-generated completeness certificate.

\begin{table}[t]
\centering\small
\caption{Coverage, accuracy, and unsupported-answer counts on
frozen CollegeMsg (282 task runs). Unsupported-answer rates are not
measured for baselines without typed trace citations.}\vspace{-2ex}
\label{tab:coverage}
\setlength{\tabcolsep}{2.5pt}
\begin{tabular}{@{}lcccc@{}}
\toprule
 & coverage & \begin{tabular}{@{}c@{}}overall\\accuracy\end{tabular}
 & \begin{tabular}{@{}c@{}}conditional\\accuracy\end{tabular}
 & \begin{tabular}{@{}c@{}}unsupported\\answers\end{tabular} \\
\midrule
TGMS, gated    & 0.706\;(199/282) & 0.408 & 0.548 & 0/199 \\
TGMS, ungated  & 0.780\;(220/282) & 0.418 & 0.536 & 21/220 \\
bi-temp.\ SQL  & 0.943\;(266/282) & 0.284 & 0.301 & n/a \\
text-to-Cypher & 0.865\;(244/282) & 0.152 & 0.176 & n/a \\
graph RAG      & 0.592\;(167/282) & 0.064 & 0.108 & n/a \\
vector RAG     & 0.486\;(137/282) & 0.106 & 0.219 & n/a \\
\bottomrule
\end{tabular}
\end{table}

\researchquestion{rq:cost}{
What do the contracts cost, and how does the architecture scale?
}

The contracts add little direct overhead. Plan admission costs
4.1\,ms at p50, and claim verification costs 3.3\,ms, together
well below 0.01\% of end-to-end task time. The 1.6\,s difference between
the gated and ungated runs in Table~\ref{tab:cost} includes model
variation and should not be interpreted as verifier latency.

TGMS keeps prompt size independent of raw graph size because it does
not serialize the store into the model context. Prompt use remains
between 6{,}000 and 10{,}000 tokens per task across the tested scales.
Database latency grows with the store but remains subsecond at
$10^7$ events: p50 latency reaches 272\,ms for a two-hop snapshot and
485\,ms for a global diff. Plans that exceed the resource budget are
rejected before execution.

The cross-pipeline differences in Table~\ref{tab:cost} are therefore
dominated by prompting, generation, and database work rather than
contract enforcement. Across model sizes, quantization settings, and
model families, accuracy ranges from 0.138 to 0.628. The gate enforces
the same support boundary across these configurations, while task
accuracy continues to depend on the underlying model.

\begin{table}[t]
\centering\small
\caption{Per-task cost on frozen CollegeMsg (282 task runs per
system). Contract-check latencies are measured separately over 108
oracle plans.}\vspace{-2ex}
\label{tab:cost}
\begin{tabular}{@{}lccccc@{}}
\toprule
 & \begin{tabular}{@{}c@{}}TGMS\\gated\end{tabular}
 & \begin{tabular}{@{}c@{}}TGMS\\ungated\end{tabular}
 & \begin{tabular}{@{}c@{}}text-to-\\Cypher\end{tabular}
 & \begin{tabular}{@{}c@{}}graph\\RAG\end{tabular}
 & \begin{tabular}{@{}c@{}}vector\\RAG\end{tabular} \\
\midrule
median wall-clock (s) & 114.1 & 112.5 & 132.6 & \textbf{14.4} & 154.0 \\
mean LLM tokens & 8{,}218 & 8{,}200 & 3{,}011 & \textbf{445} & 20{,}065 \\
\bottomrule
\end{tabular}
\end{table}

\researchquestion{rq:coverage}{
What can the current interface express on independently written
questions?
}

Four students independently authored 110 questions using only
natural-language descriptions of the datasets and the two-clock model.
They did not see the operator list, repository, or benchmark tasks.  Only 10 questions
were expressible with the current algebra. The dominant missing
capability was grouped or distinct aggregation, followed by additional
arithmetic, property-filtered patterns, and calendar semantics. In 30
questions, grouped or distinct aggregation was the only missing
operation. The study also revealed a documentation gap: users need the
interface to expose the dataset's temporal coverage and explain how
transaction time relates to ingest and correction history.

Only five expressible questions had answer types supported by the
current scorer. Across three seeds, these produced nine CollegeMsg and
six Bitcoin-OTC runs. Each system scored 3/9 on CollegeMsg and 0/6 on Bitcoin-OTC. On Bitcoin-OTC, TGMS
abstained, whereas the query-generation baselines emitted incorrect
answers. The study is small, but it clarifies the current interface boundary.
The algebra supports entity-anchored temporal traversal and
time series, while relational aggregation is the most prominent
coverage gap in this sample.

\noindent\textbf{Discussion and current scope.}
The program-generated benchmark is co-designed with the operator
interface, so its accuracy results apply to questions expressible by
the current algebra. The independent study provides a separate
coverage estimate and shows that many naturally written questions
require capabilities the algebra does not yet provide. The experiments
target different parts of the architecture: the same-information
SQL baseline compares interface designs, the result-field ablation
tests a plan-contract mechanism, and the completeness ablation tests an
evidence-contract mechanism. Together, they evaluate plan
admissibility, execution behavior, and claim faithfulness on the
studied workloads, not correctness over arbitrary temporal-graph
questions.

The guarantee envelope remains limited to the claim types currently
gated; temporal-pattern claims are checked but not yet withheld. The
verifier is implemented and mutation-tested rather than formally
verified. Because the bi-temporal SQL baseline does not implement the
evidence contract, the evaluation does not establish that claim
checking requires a fixed operator algebra. Approximate operators and
agent write-back are also outside the current prototype. These
boundaries motivate the broader research agenda in
Section~\ref{sec:agenda}.

\section{Positioning}
\label{sec:related}

TGMS intersects work on agent planning, controlled tool execution,
provenance, and temporal memory, but places the primary abstraction at
the boundary between execution and user-facing claims.

\emph{Agent planning and interface design.} Agent-first system
designs~\cite{liu2026agentfirst} redesign data systems for the scale,
redundancy, and steerability of agentic workloads.
CAESURA~\cite{urban2024caesura} and AOP~\cite{wang2025aop} ask how an
LLM can construct or orchestrate executable plans.
TAG~\cite{biswal2025tag} broadens the query classes systems should
answer. BridgeScope~\cite{weng2026bridgescope} aligns fine-grained
database tools with privileges and efficient tool-to-tool data
movement. These systems ask how agents should formulate and execute
data operations; TGMS focuses on the enforceable boundary around that
execution.

\emph{Policy, traces, and verification.} Data flow
control~\cite{summers2026dfc} enforces which flows are permitted; the
evidence contract defines the conditions under which a result supports
a claim.
Tool-execution contracts~\cite{liu2026toolgate} share the plan-side
motivation. Deep-research runtimes~\cite{russo2026deepresearch} similarly
require a support semantics for citations that refer to intermediate
data operations. We are not aware of a prior agent-database system that
combines static result-field checking, completeness propagation,
transaction-time evidence identity, and claim gating against
deterministic result digests.

\emph{Temporal memory and classical foundations.} Temporal agent
memories~\cite{zep2025,wang2026toki} adopt bi-temporal storage for
recall; TGMS uses transaction time as a component of execution
evidence (Example~A). Provenance
semirings~\cite{green2007provenance} provide a formal foundation for
lineage; TGMS adds execution conditions that lineage alone does not
capture, including belief state and result completeness. ReAct-style
loops~\cite{yao2023react} interleave planning and execution; a
whole-plan contract can inspect cost and dataflow before anything
runs, which per-call schemas cannot.

\section{Research agenda for agent-facing data systems}
\label{sec:agenda}

TGMS represents one point in a broader design space. Its limitations
suggest research problems that extend beyond temporal graphs to
relational, document, streaming, multimodal, and federated systems.

\noindent\textbf{From admission to negotiation.}
A plan contract need not stop at acceptance or rejection. Once a
complete plan is visible, the system can rewrite subplans, push
predicates, share computation, select cheaper operators, or propose an
approximate alternative under a declared error budget. This creates a
new optimization setting in which the agent and the data system
negotiate over a machine-checkable plan.

\noindent\textbf{Evidence algebras.}
Results may be incomplete, stale, approximate, or tied to a particular
belief state. These properties must propagate through joins,
aggregations, graph operators, and multi-step plans. A general evidence
algebra should define how execution conditions compose and how they
limit downstream claims. This extends provenance from where a result
came from to when and under what conditions it supports a
conclusion~\cite{green2007provenance}.

\noindent\textbf{Claim-aware proof obligations.}
Different claims require different evidence. Membership claims need
supporting tuples or entities, while complete-set claims also require
evidence that no valid member was omitted. Top-$k$, extremal, and
negative claims may require ranking guarantees, global checks, or
database-generated certificates. A system should certify a claim only
when it can discharge the corresponding proof obligation.

\noindent\textbf{Workload-driven interface design.}
A compact operator algebra simplifies planning and verification but
limits coverage. Our independent study identifies grouped aggregation,
additional arithmetic, property-filtered patterns, and calendar
semantics as important gaps. More broadly, agent-facing interfaces
should be designed from representative workloads by measuring
expressibility, marginal coverage, and the tradeoff between a small
planning space and a broad query surface.

\noindent\textbf{Belief states and cross-system evidence.}
Reproducibility over corrected data requires the system to record which
state each computation read. With concurrent agents, these states must
be selected, isolated, versioned, and audited. The same evidence must
also remain meaningful across relational queries, graph operations,
vector search, streams, and external tools, even when no single system
controls the entire plan.

Traditional database interfaces assume that clients formulate
appropriate operations and interpret results correctly. LLM agents do
not reliably satisfy either assumption. The two contracts explored in
TGMS suggest a broader role for data systems: not to make agents
infallible, but to make their plans admissible, their execution
conditions explicit, and their claim support auditable.

\bibliographystyle{ACM-Reference-Format}
\bibliography{references}

\end{document}